\documentclass[12pt]{article}
\usepackage{epsfig}
\newcommand{\bsg}{\mbox{$b \rightarrow s \gamma \ $}}
\newcommand{\bsgb}{\mbox{$\bar{b} \rightarrow \bar{s} \gamma \ $}}
\newcommand{\tz}{\mbox{$\tan \beta \ $}}

\newcommand{\tm}{\mbox{$\tilde m$}}

\newcommand{\dLRbs}{\mbox{$\delta^{LR}_{23} \ $}}
\newcommand{\dLLbs}{\mbox{$\delta^{LL}_{23} \ $}}
\newcommand{\dRLbs}{\mbox{$\delta^{RL}_{23} \ $}}
\newcommand{\dRRbs}{\mbox{$\delta^{RR}_{23} \ $}}
\newcommand{\nn}{\nonumber}
\newcommand{\raw}{\rightarrow}
\newcommand{\lraw}{\leftrightarrow}
\newcommand{\be}{\begin{equation}}
\newcommand{\ee}{\end{equation}}
\newcommand{\bea}{\begin{eqnarray}}
\newcommand{\eea}{\end{eqnarray}}

\begin{document}
%
%
\vskip -1cm
\begin{flushright}
{CERN-TH/2001-354}\\
{MCTP-01-58} \\
{\tt hep-ph/0112126}
\end{flushright}
\vspace*{0.25cm}
%

\begin{center}


{\Large{\bf Alternative approach to \bsg in the uMSSM} }

\vspace{1cm}
{\large L. Everett$^{1,2}$, G.L. Kane$^1$, S. Rigolin$^1$, 
Lian-Tao Wang$^1$ and Ting~T.~Wang$^1$} \\
\vspace{0.15cm}
{\it \footnotesize $^1$ Michigan Center for Theoretical Physics, 
University of Michigan, Ann Arbor, MI-48109 \\
$^2$ Theoretical Physics Division, CERN, CH-1211, Geneve 23, 
Switzerland}
%
%
\vskip 1cm 
\begin{abstract} 
\noindent 
The gluino contributions to the $C'_{7,8}$ Wilson coefficients for 
\bsg are calculated within the unconstrained MSSM. New stringent 
bounds on the \dRLbs and \dRRbs mass insertion parameters are 
obtained in the limit in which the SM and SUSY contributions to 
$C_{7,8}$ approximately cancel. Such a cancellation can plausibly 
appear within several classes of SUSY breaking models in which the 
trilinear couplings exhibit a factorized structure proportional to 
the Yukawa matrices. Assuming this cancellation takes place, we perform 
an analysis of the \bsg decay. We show that in a supersymmetric world 
such an alternative is reasonable and it is possible to saturate the 
\bsg branching ratio and produce a CP asymmetry of up to $20\%$, from 
only the gluino contribution to $C'_{7,8}$ coefficients. Using photon 
polarization a LR asymmetry can be defined that in principle allows for 
the $C_{7,8}$ and $C'_{7,8}$ contributions to the \bsg decay to be 
disentangled. In this scenario no constraints on the ``sign of $\mu$'' 
can be derived. 
\end{abstract} 
%
\end{center} %
\section{Introduction}
%
The precision measurements of the inclusive radiative decay $B \raw X_s 
\gamma$ provides an important benchmark for the Standard Model (SM) and 
New Physics (NP) models at the weak-scale, such as low-energy 
supersymmetric (SUSY) models. In the SM, flavor changing neutral 
currents (FCNC) are forbidden at tree level. The first SM contribution to 
the \bsg transition appears at one loop level due to the CKM flavor 
changing structure, showing the characteristic Cabibbo suppression. 
NP contributions to \bsg typically also arise at one loop, and in 
general can be  much larger than the SM contributions if no mechanisms 
for suppressing the new sources of flavor violation exist. 

Experimentally, the inclusive $B \raw X_s \gamma$ branching ratio has 
been measured by ALEPH \cite{aleph}, BELLE \cite{belle} and CLEO 
\cite{cleo} resulting in the current experimental weighted average 
\begin{equation}
BR(B\rightarrow X_s \gamma)_{exp}=(3.23 \pm 0.41) \times 10^{-4},
\label{bsgex}
\end{equation} 
with new results expected shortly from BABAR and BELLE which could further
reduce the experimental errors. Squeezing the theoretical uncertainties
down to the 10\% level has been (and still is) a crucial task. The SM
theoretical prediction has been the subject of intensive theoretical
investigation in the past several years. From the original calculation at 
LO \cite{LO}, impressive progress in the theoretical precision has been
achieved with the completion of NLO QCD calculations
\cite{NLO,misiak,degrassi1} and the addition of several further refinements 
\cite{kagan,NLOmore}.  The original complete SM NLO calculation
\cite{misiak} gives the following prediction for $\sqrt{z}=m_c/m_b=0.29$:
\begin{equation} 
BR(B\rightarrow X_s \gamma)_{SM} = (3.28 \pm 0.33) \times 10^{-4}. 
\end{equation} 
The main source of uncertainty of the previous result is due to NNLO QCD 
ambiguities. In \cite{gambino} it is shown that using $\sqrt{z} = 0.22$ 
(i.e. the running charm mass instead of the pole mass) is more justifiable and 
causes an enhancement of about 10\% of the \bsg branching ratio, 
leading to the current preferred value:
\begin{equation} 
BR(B\rightarrow X_s \gamma)_{SM} = (3.73 \pm 0.30) \times 10^{-4}.
\end{equation} 
Although these theoretical uncertainties can be addressed
only with a complete NNLO calculation, the SM value for the branching 
ratio is in agreement with the experimental measurement within the 
$1-2\sigma$ level.

The general agreement between the SM theoretical prediction and the 
experimental results have provided useful guidelines for constraining 
the parameter space of models with NP present at the electroweak 
scale, such as the 2HDM and the minimal supersymmetric standard model 
(MSSM). In SUSY models superpartners and charged Higgs loops contribute 
to \bsg$\!$, with contributions that typically rival the SM one in size. 
To get a sense of the typical magnitudes of the SUSY contribution 
to \bsg$\!$, it is illustrative to consider the (unphysical) limit of 
unbroken SUSY but broken electroweak gauge symmetry, which corresponds 
to the supersymmetric Higgsino mass parameter $\mu$ set to zero, and the 
ratio of Higgs vacuum expectation values \tz$\equiv v_u/v_d$ set to 1. 
In this limit SM and SUSY contributions are identical in size and 
cancel each other \cite{ferrara}, due to the usual sign difference 
between boson and fermion loops. Of course, this limit is unphysical: 
not only must SUSY be (softly) broken, but $\mu=0$ and \tz$=1$ have 
been ruled out by direct and indirect searches at LEP. 

In the realistic case of softly broken SUSY, the contributions to
\bsg depend strongly on the parameters of the SSB Lagrangian, as well 
as the values of $\mu$ and $\tan \beta$. In particular, as the origin 
and dynamical mechanism of SUSY breaking are unknown, there is no 
reason {\it a priori} to expect that the soft parameters will be 
flavor-blind (or violate flavor in the same way as the SM). Of course, 
the kaon system has provided strong FCNC constraints for the mixing of the 
first and second generations which severely limit the possibility of 
flavor violation in that sector \cite{gabbiani,kaon}. Note however 
that the constraints for third generation mixings are significantly weaker, 
with \bsg providing usually the most stringent constraints.

Nevertheless, for calculational ease one of the following simplified 
MSSM scenarios have often been assumed:
\begin{itemize}
\item 
The SUSY partners are very heavy and their contribution decouples, 
so that only the Higgs sector contributes to \bsg$\!$. In this scenario, as 
well in general 2HDMs, NLO calculations have been performed 
\cite{degrassi1,ciafaloni,borzumati2h}. Due to coherent contributions 
between SM and Higgs sector, a lower bound on the charged Higgs mass can
usually be derived \cite{deboer} in this class of models. In the large 
\tz region the two-loop SUSY correction to the Higgs vertex can produce 
quite sizeable modifications and should be carefully taken into 
account \cite{degrassi3}.  
\item 
The SUSY partners as well the extra Higgs bosons have masses of order 
the electroweak scale, but the only source of flavor violation is in the 
CKM matrix.  This scenario, known as minimal flavor violation (MFV), is 
motivated for example within minimal supergravity (mSUGRA) models. MFV 
scenarios have been studied at LO \cite{bertolini,mfv,baer,masiero},  
in certain limits at NLO \cite{degrassi2}, and including large \tz 
enhanced two-loop SUSY contributions \cite{degrassi3,demir}. 
In this scenario, the \bsg decay receives a contribution 
from the chargino sector as well as from the charged Higgs sector.
To avoid overproducing \bsg $\!$, the charged Higgs and chargino loops
must cancel to a good degree.  This cancellation can be achieved for a
particular ``sign of $\mu$'' in the mSUGRA parameter space\footnote{
Specifically the relative sign between the parameters $\mu$ and $A_t$, 
and so generally different from the ``sign of $\mu$'' relevant in 
the case of the muon $g-2$ MSSM contribution \cite{everett}.} which 
flips the sign of the chargino contribution relative to the SM and 
charged Higgs loops, always interfering constructively. Although 
this cancellation can occur and puts important constraints on the 
mSUGRA parameter space, it is important to note that it is not due to 
any known symmetry but rather should be interpreted, in an certain 
sense, as a fine-tuning.
\item 
There are new sources of flavor violation in the soft breaking terms.  In
this case, additional SUSY loops involving down-type squarks and gluinos 
or neutralinos (hereafter neglected compared with the gluino loops due to 
the weaker coupling) contribute to \bsg$\!$. It is well known that the 
gluino contribution can dominate the amplitude for such nonminimal SUSY 
models, both due to the $\alpha_s/\alpha$ enhancement with respect to the 
other SM and SUSY contributions, and due to the $m_{\tilde{g}}/m_b$ 
enhancement from the chirality flip along the gluino line.  Thus in this 
scenario, which is generally noted as the unconstrained MSSM (uMSSM),
usually only the gluino contribution is discussed. It has been shown
\cite{gabbiani,pokorski} that the 23-LR off-diagonal entry of the
down-squark mass matrix is severely constrained by \bsg measurements 
to be of ${\cal O}(10^{-2})$. Less stringent bounds can be obtained for 
the other 23 off-diagonal entries. No known symmetry assures that these 
constraints can be automatically satisified; again this fact could be 
interpreted at the electroweak scale as a fine-tuning. 
\end{itemize}

\noindent
A discussion of the \bsg process in the general unconstrained MSSM is in 
principle possible, but it is necessary to deal with two unavoidable 
problems: (i) a large number of free, essentially unconstrained 
parameters, and (ii) the need to achieve a quite accurate cancellation 
between the sizeable different contributions (SM, Higgs, 
chargino/neutralino and gluino) to the Wilson coefficient $C_7$ 
associated with the $Q_7 \propto m_b \bar{s}_L \sigma^{\mu\nu} b_R 
F_{\mu \nu}$ operator in such a way that the experimental measurement, 
which approximately saturated solely by the SM result, is satisfied. 
Moreover, in general MSSM models with nonminimal flavor violation the 
gluino loop can also contribute significantly to the Wilson coefficient 
$C'_7$ associated with  the chirality-flipped operator, $Q'_7 \propto 
m_b \bar{s}_R  \sigma^{\mu\nu} b_L F_{\mu \nu}$, as has been recently 
emphasized in the literature \cite{borzumati,besmer}. However, as the SM, 
Higgs, and chargino contributions to $C'_7$ are typically suppressed by 
a factor of $O(m_s/m_b)$, it is not possible in general to achieve a 
cancellation between the different terms in $C'_7$ and thus a stronger 
fine-tuning has to be imposed. 

However, it has been recently shown \cite{kv} that in many classes of 
SUSY breaking models a particular structure of the soft trilinear 
couplings $\tilde{A}$ of the soft-breaking Lagrangian 
can be derived which can alleviate these constraints.
Writing these couplings as $\tilde{A}_{ij}=A_{ij} Y_{ij}$ (in which $Y$ 
denotes the fermion Yukawa matrices), the matrices $A$ for the up and 
down sector are given respectively by:
\be
A^{(u)}_{ij} = A^L_{ii} + A^{R,u}_{jj} \qquad , \qquad  
A^{(d)}_{ij} = A^L_{ii} + A^{R,d}_{jj}~.  
\label{tril}
\ee
As shown in \cite{kv}, this factorization holds quite generally in string 
models, for example in Calabi-Yau 
models in the large $T$ limit or in Type I models \cite{imr}, as well as 
in gauge-mediated \cite{giudice1} and anomaly-mediated models 
\cite{randall,giudice2,rattazzi}. If eq.(\ref{tril}) holds, 
specific relations can be derived for the off-diagonal LR entries in 
squark mass matrix. In particular, the leading contribution to the 
entries of interest for the \bsg process are given in the SCKM basis as:
\bea
\tilde{A}^{(u)}_{23} & \propto & m_t \left[ 
  (A^L_{22}-A^L_{11}) (V^{(u)}_L)_{22} (V^{(u)}_L)^*_{32} +  
  (A^L_{33}-A^L_{11}) (V^{(u)}_L)_{23} (V^{(u)}_L)^*_{33} \right]~, 
  \label{At23} \\
\tilde{A}^{(u)}_{32} & \propto & m_t \left[ 
  (A^{R,u}_{22}-A^{R,u}_{11}) (V^{(u)}_R)_{32} (V^{(u)}_R)^*_{22} +  
  (A^{R,u}_{33}-A^{R,u}_{11}) (V^{(u)}_R)_{33} (V^{(u)}_R)^*_{23} \right]~, 
  \label{At32} \\  
\tilde{A}^{(d)}_{23} & \propto & m_b \left[ 
  (A^L_{22}-A^L_{11}) (V^{(d)}_L)_{22} (V^{(d)}_L)^*_{32} +  
  (A^L_{33}-A^L_{11}) (V^{(d)}_L)_{23} (V^{(d)}_L)^*_{33} \right]~, 
  \label{Ab23}\\
\tilde{A}^{(d)}_{32} & \propto & m_b \left[ 
  (A^{R,d}_{22}-A^{R,d}_{11}) (V^{(d)}_R)_{32} (V^{(d)}_R)^*_{22} +  
  (A^{R,d}_{33}-A^{R,d}_{11}) (V^{(d)}_R)_{33} (V^{(d)}_R)^*_{23} \right]~, 
  \label{Ab32} 
\eea
with $V^{(u,d)}_{L,R}$ the rotation matrices for the up and down quark 
sector from the interaction to the mass eigenstate\footnote{In this 
notation the CKM matrix is $V_{CKM} = V^{(u)}_L (V^{(d)}_L)^\dagger$.}. 
From eqs.(\ref{At23}-\ref{Ab32}), one can realize first that the 
down-sector LR off-diagonal entries are naturally suppressed by a factor 
of $O(m_b/m_t)$ compared with the up-squark sector ones
due to the  particular factorization of the soft trilinear couplings 
given in eq.(\ref{tril}). Second, in these classes of models both the $23$ 
and $32$ entries are of the same order and proportional to the largest 
mass (up or down). 
Consequently, in these classes of models, ${\cal O}(10^{-2})$ 
off-diagonal entries in the down-squark sector along with ${\cal O}(1)$ 
off-diagonal entries in the up-squark can be considered in some sense as a 
prediction of the underlying fundamental theory\footnote{It is important
to note however that the off-diagonal entries of $\tilde{A}$ in the SCKM
basis contain terms proportional to the products of entries of the
left-handed and right-handed quark rotation matrices, which are largely
unconstrained (except for the CKM constraint for the left-handed up and
down quark rotation matrices which enter (for example) $\tilde{A}_{23}$).
The quark rotation matrices are highly model-dependent. While the
diagonal entries can in general safely to be taken ${\cal O}(1)$, it is
typically assumed that the off-diagonal quark rotation matrices are 
suppressed by powers of the Cabibbo angle in a way that mirrors the CKM
matrix (see e.g. \cite{kv}). Note though that this assumption is not
required, particularly for the right-handed quark rotation matrices
which enter  $\tilde{A}_{32}$ which are of particular relevance for this 
paper.}. 
This fact implies comparable chargino and gluino contributions 
to \bsg$\!$, making the possibility of cancellations between the W and the 
different SUSY contributions to the $Q_7$ operator less unnatural. 
The constraints on the gluino contribution to $Q'_7$ are simultaneously  
alleviated. This flavor structure holds in essentially all attempts to build 
string-motivated models of the soft-breaking Lagrangian.

The structure of the paper is as follows. In section \ref{sectionb}, we
briefly summarize the theoretical framework for the calculation of the 
\bsg branching ratio at LO and NLO. In section \ref{sectionc}, we derive 
useful mass insertion (MI) formulas for the gluino contributions to the 
Wilson coefficients $C_{7,8}$ and $C'_{7,8}$.
We demonstrate explicitly that in the large 
\tz region, a good understanding of these expressions is obtained 
only by retaining terms in the MI expansion through the second order. 
For $\mu$ of the same order as the common squark mass parameter and 
large $\tan \beta$, new (previously overlooked) off-diagonal terms become 
relevant in the \bsg process. 
We then devote our attention in section \ref{sectiond} 
to the analysis of the gluino contribution to $C'_{7,8}$ in the general 
uMSSM. In particular, we ask the question of whether the contribution to 
$C'_7$ alone can saturate the \bsg branching ratio, assuming that the SM 
and SUSY contributions to $C_7$ cancel each other to an extent that the 
effects of $C_7$ are subleading. While this scenario may initially appear 
to be unnatural, we will argue that sufficient cancellations in $C_7$ do 
not involve significantly more fine tuning than the usual cancellation 
required in MFV scenarios. With this analysis, we thus provide an alternative 
interpretation of \bsg which is at least as viable as any supersymmetric 
one. This analysis also provides more general mass insertion bounds on 
\dRLbs than those obtained recently \cite{besmer}, where the SM (and 
sometimes Higgs and chargino) contributions to $C_7$ are always retained.  
As we are generally interested in moderate to  large values of \tz, we are 
able to put rather stringent bounds on the mass insertion parameter 
\dRRbs$\!$. In section \ref{sectiondc}, we study the branching 
ratio and CP asymmetry as functions of the SUSY parameter space within 
this scenario, assuming complex off-diagonal MIs. Throughout the paper, 
to avoid EDM constraints we set the relevant reparameterization 
invariant combinations of the flavor-independent phases 
to zero. Finally in section 
\ref{sectiondd} we show that if the photon polarization 
will be measured, it is possible to distinguish such a scenario from the 
usual $C_7$ dominated scenario through the definition of a LR asymmetry.

Since we are interested in analyzing a supersymmetric world where 
the one-loop SUSY effects are of the same order as the SM loops, we 
assume relatively light superpartner masses. Specifically we choose the 
gluino mass $\tm_{\tilde{g}}=350$ GeV and the common diagonal down-squark 
mass $\tm_D=500$ GeV, with the lightest down-squark mass in the $250-500$ 
GeV range. All of the other sfermion masses, as well the chargino and 
neutralino masses, do not enter directly in our analysis and (some of 
them) can be taken to be reasonably light as suggested by 
\cite{altarelli}. Motivated by the lower limit on the Higgs boson mass 
\cite{janot} (which suggests $|\cos 2 \beta| \approx 1$) and by the muon 
$g-2$ excess, we focus to some extent on moderate to large values of $\tan 
\beta$, though our formulas and much of the analysis hold in general. 
%
\section{\bsg branching ratio at NLO}
\label{sectionb}
%
For the purpose of presentation, we summarize the theoretical 
framework for evaluating the \bsg branching ratio at NLO. A complete and 
detailed discussion can be found for example in 
\cite{misiak,degrassi1,kagan}. 
The starting point in the calculation of the B meson decay rates is the 
low-energy effective Hamiltonian, at the bottom mass scale $\mu_b$:
\be 
{\cal H}_{eff} = - \frac{4 G_F}{\sqrt{2}} V_{tb} V^*_{ts} 
   \sum_i C_i(\mu_b) Q_i(\mu_b) \ .  
\ee 
The operators relevant to the \bsg process\footnote{This of course depends 
on the basis chosen; we have chosen the one easiest for our 
discussion.} are: 
\bea 
Q_2 & = & \bar{s}_L \gamma_\mu c_L \bar{c}_L \gamma^\mu b_L \ , \nn \\
Q_7 & = & \frac{e}{16 \pi^2} m_b \bar{s}_L \sigma^{\mu \nu} b_R F_{\mu \nu} \ , 
        \nn \\
Q_8 & = & \frac{g_s}{16\pi^2} m_b \bar{s}_L \sigma^{\mu \nu} G^a_{\mu \nu} 
        T_a b_R \ . \label{opQ}
\eea 
and their $L \lraw R$ chirality counterpart:
\bea 
Q'_2 & = & \bar{s}_R \gamma_\mu c_R \bar{c}_R \gamma^\mu b_R \ , \nn \\ 
Q'_7 & = &\frac{e}{16\pi^2} m_b \bar{s}_R \sigma^{\mu \nu} b_L F_{\mu \nu} \ , 
         \nn \\ 
Q'_8 & = &\frac{e}{16\pi^2} m_b \bar{s}_R \sigma^{\mu \nu} G^a_{\mu \nu} 
         T_a b_L \ . \label{opQp}
\eea 

The Wilson coefficients $C^{ (_{'} ) }_{2,7,8}$ are initially evaluated
at the electroweak or soft SUSY breaking scale, which we generically 
denote as $\mu_0$, and then evolved down to the bottom mass scale $\mu_b$. 
The standard\footnote{In a recent paper \cite{borzumati} it has been 
pointed out that the gluino contribution (and the same argument holds 
also for the chargino and neutralino contributions) is the sum of two 
different pieces, one proportional to the bottom mass and one proportional 
to the gluino mass, which have a different RG evolution. We have found 
that at LO, this is equivalent to the usual SM evolution once the running 
bottom mass $m_b(\mu_0)$ is used instead of the pole mass in the 
$C_i(\mu_0)$ WC.} RG equations for the $C_{2,7,8}$ operators from the 
electroweak scale ($\mu_W < m_t$) to the low-energy scale $\mu_b$ 
is given by:
\bea
C_2 (\mu_b) &=& \frac{1}{2}\left(\eta^{-\frac{12}{23}} + 
   \eta^{\frac{6}{23}} \right) C_2(\mu_W) \ , \label{C2mub} \\
C_7 (\mu_b) &=& \eta^{\frac{16}{23}} C_7(\mu_W) + \frac{8}{3}
   \left( \eta^{\frac{14}{23}} -\eta^{\frac{16}{23}} \right) C_8(\mu_W) + 
   \sum_{i=1}^8 h_i \eta^{a_i} \ , \label{C7mub} \\
C_8 (\mu_b) &=& \eta^{\frac{14}{23}} C_8(\mu_W) + 
   \sum_{i=1}^8 \bar{h}_i \eta^{a_i} \ , \label{C8mub} 
\eea
where $\eta=\alpha_s(\mu_W)/\alpha_s(\mu_b)$ and $h_i,\bar{h}_i$ and 
$a_i$ are constants (see \cite{misiak} for details). The $C'_{2,7,8}$ 
coefficients obey the same running as their chirality conjugate 
counterparts. If the NP scale is much higher than $m_t$, the running from 
$\mu_{SUSY}$ to $\mu_W$ with six quarks should also be 
taken into account (see the first paper of \cite{degrassi3}).  
The coefficient $C_2$ is dominated by a SM tree-level diagram and is 
normalized such that $C_2(\mu_W) = 1$. Its chirality conjugate, $C'_2$, 
has no SM contribution at tree level and can thus be safely set to 
zero. The NP contributions to $C_2$ and $C'_2$ appear at one-loop 
order and are negligible. The Wilson coefficients $C_7$ and $C'_7$ are 
the only coefficients that contribute directly to the \bsg branching 
ratio at the lowest QCD order ($\alpha_s^0$). These coefficients receive 
contributions both from the SM and NP at one-loop order. The coefficients 
$C_8$ and $C'_8$ receive one-loop SM and NP contributions through the 
same types of diagrams as $C_7$ and $C'_7$, but with the external photon 
line substituted by a gluon line. When the QCD running from the matching 
scale $\mu_0$ to $\mu_b$ is performed, these different coefficients mix, 
as shown in eqs.(\ref{C2mub}-\ref{C8mub}), so that the ``effective'' 
low-energy coefficients $C_{2,7,8}(\mu_b)$ receives contributions from 
different operators.
 
The \bsg branching ratio is usually defined by normalizing it to the 
semileptonic $b \raw c \ e^- \ \bar{\nu}_e$ branching ratio, giving:
\be
BR(B \raw X_s \gamma)|_{E_\gamma > (1-\delta) E^{max}_\gamma} = 
BR(B \raw X_c e \bar{\nu}) \frac{6 \alpha}{\pi f(z)} 
\left| \frac{V_{tb} V_{ts}^*}{V_{cb}} \right|^2 K(\delta,z)~.
\ee
Here $f(z)$ is a phase space function and should be calculated for 
on-shell masses, namely $\sqrt{z} = m_c/m_b = 0.29$. $\delta$ is the 
experimental photon detection threshold, which for comparison between 
experimental data and theoretical prediction is usually set to $0.9$ 
\cite{kagan}. 
The dependence of $K_{NLO}$ from the Wilson coefficients $C_i$ and 
$C'_i$ at NLO is given by \cite{kagan}:
\bea
K_{NLO}(\delta,z) = & \sum_{i \leq j=2,7,8} & k^{(0)}_{ij}(\delta,z) \ 
   \left\{ {\rm Re}[C^{(0)}_i(\mu_b) C^{(0)*}_j(\mu_b)] \ +  
   \left( C_{i,j} \raw C'_{i,j} \right) \right\} \ + \nn \\ 
& & k^{(1)}_{77}(\delta,z) \left\{ {\rm Re}[C^{(1)}_7(\mu_b) 
   C^{(0)*}_7(\mu_b)] \ + \left( C_7 \raw C'_7 \right) \right\}~.
\label{knlo}
\eea
In the previous expression $C^{(0)}_i$ and $C^{(1)}_i$ refer respectively 
to the LO and NLO contributions to the Wilson coefficients $C_i$ defined 
as:
\be
C_i (\mu_b) = C_i^{(0)}(\mu_b) + \frac{\alpha_s(\mu_b)}{4\pi} 
   C^{(1)}_i (\mu_b) + {\cal O}(\alpha, \alpha_s^2)~. 
\ee
As in the following we are deriving only one-loop formulas for the 
Wilson coefficients $C^{(_{'})}_{7,8}$, $C_i \equiv C_i^{(0)}$. 
We will briefly discuss the effects of including $C^{(1)}_7$ in 
section \ref{sectionda}. The coefficients $k_{ij}(\delta,z)$ used in the 
our analysis are calculated for $\delta=0.9$ and $\sqrt{z} = 0.22$ using the 
formulas derived in \cite{misiak,kagan}.
The LO branching ratio expression can be easily derived from eq.(\ref{knlo}) 
setting $k^{(0)}_{77}=1$ and all the other $k^{(0,1)}_{ij}=0$, giving:
\be
K_{LO} = |C_7(\mu_b)|^2 + |C'_7(\mu_b)|^2~,
\ee  
independently of the choice of $\delta$ and $z$.
%
\section{$C_{7,8}$ and $C'_{7,8}$ gluino contributions to \bsg}
\label{sectionc}
%
In the following we will focus on the gluino contribution to the 
Wilson coefficients $C_{7,8}$ and $C'_{7,8}$. There is only one gluino 
diagram that contributes to $C_{7}$ and $C'_7$, with the external photon 
line attached to the down-squark line, while two diagrams can contribute 
to the $C_8$ and $C'_8$ coefficients, as the gluon external line can be 
attached to the squark or the gluino lines. The one-loop gluino 
contributions to the $C_{7,8}$ and $C'_{7,8}$ coefficients are given 
respectively by:
\bea
C^{\tilde{g}}_7(\mu_W) &=&\frac{4g_s^2 }{3g^2}{{Q_d}\over{V_{tb} V^*_{ts}}} 
   \sum_{A} {{m_W^2}\over{\tm^2_A}} \left\{ L_{b} L_{s}^* F_2(x^g_A) + 
   \frac{\tm_{\tilde{g}}}{m_b} R_{b} L_{s}^* F_4(x^g_A) \right\} 
\label{glC7} \\
C^{\tilde{g}}_8(\mu_W) &=& -\frac{g_s^2}{6g^2} {{Q_d}\over{V_{tb} V^*_{ts}}} 
   \sum_{A} {{m_W^2}\over{\tm^2_A}} \left\{ L_{b} L_{s}^* F_{21}(x^g_A) + 
   \frac{\tm_{\tilde{g}}}{m_b} R_{b} L_{s}^* F_{43}(x^g_A) \right\} 
\label{glC8} \\
C^{' \tilde{g}}_7(\mu_W) &=&\frac{4g_s^2}{3g^2}{{Q_d}\over{V_{tb} V^*_{ts}}}
   \sum_{A} {{m_W^2}\over{\tm^2_A}}  
   \left\{ R_{b} R_{s}^* F_2(x^g_A) + 
   \frac{\tm_{\tilde{g}}}{m_b} L_{b} R_{s}^* F_4(x^g_A) \right\}
\label{glC7p} \\
C^{' \tilde{g}}_8(\mu_W) &=&-\frac{g_s^2}{6g^2}{{Q_d}\over{V_{tb} V^*_{ts}}} 
   \sum_{A} {{m_W^2}\over{\tm^2_A}} \left\{ R_{b} R_{s}^* F_{21}(x^g_A) + 
   \frac{\tm_{\tilde{g}}}{m_b} L_{b} R_{s}^* F_{43}(x^g_A) \right\}, 
\label{glC8p}
\eea
in which $x^g_A = \tm_{\tilde{g}}^2 / \tm_{\tilde{A}}^2$, with 
$\tm_{\tilde{g}}$ the gluino mass, and $\tm_{\tilde{A}}$ the mass of 
the A-th down squark eigenstate. $L_d$ and $R_d$ are the Left and 
Right gluino couplings to a generic down quark $d$ given by:
\be
L^{\tilde g}_{d} =  - \sqrt{2} \ U_{A,d} \qquad , \qquad  
R^{\tilde g}_{d} =   \sqrt{2} \ U_{A,d+3} \ , \label{glcoupl}
\ee
in which $U$ is the $6 \times 6$ down-squark rotation matrix. The loop 
integrals $F_{12}$ and $F_{43}$ are defined as:
\be
F_{21} = F_2(x) + 9 F_1(x) \qquad , \qquad  F_{43} = F_4(x) + 9 F_3(x)\ ,
\ee 
using the conventions for the integrals $F_i(x)$ as in \cite{bertolini} 
for an easier connection with the standard convention in the literature. 

It is illustrative to write the gluino contribution to the $C_{7,8}$ and 
$C'_{7,8}$ Wilson coefficients using the MI approximation. First, note 
that the set of integrals used in \cite{bertolini} is not the most 
appropriate for dealing with the MI formulas. However, for the sake of 
simplicity we will retain these conventions and further define the 
integrals $F_i$ and their ``derivatives'' through the following 
self-consistent relations:
\bea
\hskip 0.5cm
F_i(\frac{x}{y}) \equiv \frac{1}{y} f_i(x,y) \ , \quad
F^{(1)}_i(\frac{x}{y}) \equiv \frac{1}{y^2} \frac{\partial}{\partial y} 
  f_i(x,y) \ , \ ... \ , \quad
F^{(n)}_i(\frac{x}{y}) \equiv \frac{1}{n!} \frac{1}{y^{n+1}}
\frac{\partial^n}{\partial y^n} f_i(x,y)\ . & & \nn  
\eea 
Using this notation, the first and second order terms in the MI expansion 
for the $C_{7,8}$ and $C'_{7,8}$ coefficients are given respectively by:
\bea
C^{\tilde{g}}_7 (1) &=& \frac{8 g_s^2 }{3 g^2} {{Q_d}\over{V_{tb} V^*_{ts}}} 
   {{m_W^2}\over{\tm^2_D}} \left\{ \delta^{LL}_{23} F^{(1)}_2(x^g_D) - 
   \frac{\tm_{\tilde{g}}}{m_b} \delta^{LR}_{23} F^{(1)}_4(x^g_D) \right\},
\label{glC7MI} \\
C^{\tilde{g}}_8 (1) &=& - \frac{g_s^2}{3 g^2} {{Q_d}\over{V_{tb} V^*_{ts}}} 
   {{m_W^2}\over{\tm^2_D}} \left\{ \delta^{LL}_{23} F^{(1)}_{21}(x^g_D) - 
   \frac{\tm_{\tilde{g}}}{m_b} \delta^{LR}_{23} F^{(1)}_{43}(x^g_D) \right\},
\label{glC8MI} \\
C^{' \tilde{g}}_7 (1) &=& \frac{8 g_s^2}{3 g^2}{{Q_d}\over{V_{tb} V^*_{ts}}}
   {{m_W^2}\over{\tm^2_D}} \left\{ \delta^{RR}_{23} F^{(1)}_2(x^g_D) - 
   \frac{\tm_{\tilde{g}}}{m_b} \delta^{RL}_{23} F^{(1)}_4(x^g_D) \right\},
\label{glC7pMI} \\
C^{' \tilde{g}}_8 (1) &=& - \frac{g_s^2}{3 g^2} {{Q_d}\over{V_{tb} V^*_{ts}}} 
   {{m_W^2}\over{\tm^2_D}} \left\{ \delta^{RR}_{23} F^{(1)}_{21}(x^g_D) - 
  \frac{\tm_{\tilde{g}}}{m_b} \delta^{RL}_{23} F^{(1)}_{43}(x^g_D) 
\right\}, 
\label{glC8pMI}
\eea
and 
\bea
C^{\tilde{g}}_7 (2) &=& \frac{4 g_s^2 }{3 g^2} {{Q_d}\over{V_{tb} V^*_{ts}}} 
   {{m_W^2}\over{\tm^2_D}} \frac{m_b (A_b - \mu \tz)}{\tm^2_D} 
   \left\{ \delta^{LR}_{23} F^{(2)}_2(x^g_D) -
   \frac{\tm_{\tilde{g}}}{m_b} \delta^{LL}_{23} F^{(2)}_4(x^g_D) \right\},~
\label{glC7MI2} \\
C^{\tilde{g}}_8 (2) &=& - \frac{g_s^2}{6 g^2} {{Q_d}\over{V_{tb} V^*_{ts}}} 
   {{m_W^2}\over{\tm^2_D}} \frac{m_b (A_b - \mu \tz)}{\tm^2_D} 
   \left\{ \delta^{LR}_{23} F^{(2)}_{21}(x^g_D) - 
   \frac{\tm_{\tilde{g}}}{m_b} \delta^{LL}_{23} F^{(2)}_{43}(x^g_D) \right\},~
\label{glC8MI2} \\
C^{' \tilde{g}}_7 (2) &=& \frac{4 g_s^2}{3 g^2}{{Q_d}\over{V_{tb} V^*_{ts}}}
   {{m_W^2}\over{\tm^2_D}} \frac{m_b (A_b - \mu \tz)}{\tm^2_D} 
   \left\{ \delta^{RL}_{23} F^{(2)}_2(x^g_D) - 
   \frac{\tm_{\tilde{g}}}{m_b} \delta^{RR}_{23} F^{(2)}_4(x^g_D) \right\},~
\label{glC7pMI2} \\
C^{' \tilde{g}}_8 (2) &=& - \frac{g_s^2}{6 g^2} {{Q_d}\over{V_{tb} V^*_{ts}}} 
   {{m_W^2}\over{\tm^2_D}} \frac{m_b (A_b - \mu \tz)}{\tm^2_D} 
   \left\{ \delta^{RL}_{23} F^{(2)}_{21}(x^g_D) - 
   \frac{\tm_{\tilde{g}}}{m_b} \delta^{RR}_{23} F^{(2)}_{43}(x^g_D) \right\}.~ 
\label{glC8pMI2}
\eea
In the previous formulas $x^g_D=\tm^2_{\tilde{g}}/\tm^2_D$, with $\tm_D$ 
the average down-squark mass related to the down-squark mass eigenstates 
via the relation $\tm^2_A = \tm^2_D +\delta m^2_A$. The definitions of the 
MI parameters are:
\bea
\delta^{LL}_{ij} = \frac{1}{\tm^2_D} \sum_{A=1}^6 U^\dagger_{i,A} 
   \delta m^2_A U_{A,j} \quad &,& \quad
\delta^{RR}_{ij} = \frac{1}{\tm^2_D} \sum_{A=1}^6 U^\dagger_{i+3,A} 
   \delta m^2_A U_{A,j+3} \ , \quad \nn \\ 
\delta^{LR}_{ij} = \frac{1}{\tm^2_D} \sum_{A=1}^6 U^\dagger_{i,A} 
   \delta m^2_A U_{A,j+3} \quad &,& \quad 
\delta^{RL}_{ij} = \frac{1}{\tm^2_D} \sum_{A=1}^6 U^\dagger_{i+3,A} 
   \delta m^2_A U_{A,j} \ .
\label{MIdef}
\eea
In deriving eqs.(\ref{glC7MI2}-\ref{glC8pMI2}) to the second order in 
the MI parameters, we have kept only the dominant term proportional to \tz 
(the $A_b$ term is retained in the above expression for defining our 
convention for the $\mu$ term; see later), and neglected all of the other 
off-diagonal mass insertions. Clearly the dominant terms in 
eqs.~(\ref{glC7MI}-\ref{glC8pMI2}) are those proportional to the 
gluino chirality flip, such that the gluino contribution 
to $C_7$ ($C'_7$) at first order depends only on the MI term \dLRbs 
(\dRLbs$\!$). However, for large \tz and $\mu \approx \tm_A$, the second 
order MI terms in eqs.~(\ref{glC7MI2}-\ref{glC8pMI2}) can become 
comparable in size with the first order mass insertions. Thus, 
two different MI parameters are relevant in the L/R sectors: (\dLRbs$\!$, 
\dLLbs$\!$) and (\dRLbs$\!$, \dRRbs$\!$), contrary to common wisdom. To 
which extent the LL and RR MIs are relevant 
depends of course on the values chosen for $\mu$ and $\tan \beta$, but in 
a large part of the allowed SUSY parameter space they cannot in general be 
neglected. Moreover, the fact that the gluino Wilson coefficients depend 
on two different MI parameters will have important consequences in the 
study of the \bsg CP asymmetry\footnote{Specifically, if only the first 
order term in the MI is taken the \bsg CP asymmetry vanishes, as discussed 
in greater detail in section \ref{sectiondc}.}. 
%
%
\section{Alternative solution to \bsg branching ratio}
\label{sectiond}
%
%
In the majority of the previous studies of the \bsg process, the main 
focus was to calculate the SM and NP contributions to the $C_{7,8}$ 
coefficients. The contributions to \bsg coming from $C'_{7,8}$ have 
usually been neglected on the assumption that they are suppressed  
compared to $C_{7,8}$ by the ratio $m_s/m_b$. While this assumption is 
always valid for the SM and for the Higgs-sector contributions, in the 
case of the uMSSM this is not generally the case. It is only within 
specific MSSM scenarios (such as MFV) that the gluino and 
chargino contributions to the $C'_{7,8}$ coefficients can be neglected due 
to the $m_s/m_b$ suppression factor. In the general uMSSM this 
suppression can be absent and, in particular, the gluino contributions to 
$C_{7,8}$ and $C'_{7,8}$ are naturally of the same order 
\cite{borzumati,besmer}. 

Therefore, in the following we present an alternative 
approach to the \bsg process in supersymmetric models. We assume a 
particular scenario in which the {\it total} contribution to 
$C_{7,8}$ is negligible and the main 
contribution to the \bsg branching ratio is given by $C'_{7,8}$. This 
``$C'_7$ dominated'' scenario is realized when the chargino, neutralino, 
and gluino contributions to $C_{7,8}$ sum up in such a way as to cancel 
the W and Higgs contributions almost completely\footnote{The main 
constraint on this scenario is the requirements of the $C_7$ 
cancellation. The $C_8$ contribution enters in the \bsg branching 
ratio at ${\cal O} (\alpha_s)$ and usually cannot account for more 
than $10\%$ of the measured branching ratio.}. 
In our opinion this situation does not require substantially more fine 
tuning than what is required in the usual MFV scenario, where conversely 
the NP contributions to $C_{7,8}$ essentially cancel between themselves 
(or are almost decoupled) so that all the measured \bsg branching ratio is 
produced by the W diagram. As previously discussed, many classes of SUSY 
breaking models \cite{kv} lead to  off-diagonal 
LR entries of the down-squark sector that are naturally suppressed 
compared with those of the up-squark sector:
\be
(\delta^{LR}_{ij})^d \approx \frac{\max(m_i,m_j)}{m_t} (\delta^{LR}_{ij})^u
\ee
in which $m_{i,j}$ are down-quark masses. In particular, the 
$(\delta^{LR}_{23})^d$ 
entries, which are 
relevant for the $b\raw s\gamma$ process, receive a 
$O(m_b/m_t)$ 
suppression as can be derived from eqs.(\ref{At23}-\ref{Ab32}). For 
$(\delta^{LR}_{23})^u \approx {\cal{O}}(1)$, a natural value $(\dRLbs)^d 
\approx {\cal{O}}(m_b/m_t) \approx 10^{-2}$ is obtained. With this 
mechanism at work, off-diagonal chargino and gluino contributions 
to flavor changing processes are naturally of the same order. The 
$\alpha_s/\alpha_w$ enhancement of the gluino contribution with respect 
to the chargino one is compensated by the $m_b/m_t$ suppression of the LR 
off-diagonal entries.
Clearly a complete analysis of the regions of uMSSM parameter space where 
the $C_{7,8}$ cancellation takes place is an important task, 
necessary for studying the details of this scenario. However, a detailed 
analysis is beyond the scope of this paper will be discussed elsewhere 
\cite{ekrww2}. It is worth mentioning that in 
preliminary scans we checked that it is not difficult to
find a candidate set of parameters where $C_{7,8}$ numerically yield 
small contributions to the \bsg branching ratio. Of course, this 
set is not obviously not expected to be unique, and further checking 
that any such parameter sets are consistent with all the other existing 
measurements of FCNC and CP-violating observables will impose 
further strong constraints.

Finally, we stress that in the following analysis we do not 
make any specific assumptions as to the size of the off-diagonal 
entries of the down-squark mass matrix. In particular, we are not using 
any of the relations described in eqs.(\ref{tril}-\ref{Ab32}). The 
previous arguments have been intended as a theoretical framework for the 
following model independent analysis. A general discussion of the 
CP-violating sector, using the factorization ansatz of eq.(\ref{tril}), 
will be the subject of a forthcoming paper \cite{ekrww2}.

%
\subsection{Single MI dominance analysis}
\label{sectionda}
From eqs.(\ref{glC7MI}-\ref{glC8pMI2}), one can read (in MI language) 
the off-diagonal entries that are relevant for the gluino contribution 
to the $C_{7,8}$ and $C'_{7,8}$ Wilson coefficients\footnote{From now 
on, for the sake of simplicity the symbol $\delta^{AB}_{ij}$ will be used 
instead of $(\delta^{AB}_{ij})^d$ for referring to the down-squark MIs.}. 
Note that limits on \dLRbs $\approx {\cal{O}}(10^{-2})$ have previously 
been obtained in \cite{gabbiani}. No stringent bound has been derived 
there for \dLLbs, as this term at lowest order
does not come with the  $\tm_{\tilde{g}}/m_b$ enhancement (see 
eqs.(\ref{glC7MI})). No limits were derived on \dRLbs 
and \dRRbs because in the specific scenario used in \cite{gabbiani}, the 
``opposite chirality'' MIs are suppressed by a factor $m_s/m_b$ and so 
negligible. An analysis of the \dRLbs dependence has 
been performed in \cite{besmer}, in which the $W$ contribution to 
$C_{7,8}$ was not set to zero (sometimes also Higgs and MFV chargino 
contributions to  $C_{7,8}$ were included). Consequently their bounds on 
the down-squark off-diagonal MIs contributing to $C'_{7,8}$ are 
more stringent than the bounds we derive in our scenario, for which 
the total contribution to $C_{7,8}$ is assumed to be negligible. 
It is clearly only in the scenario we study that an {\em absolute} 
constraint on these MIs be derived. Moreover no analysis on \dLLbs and 
\dRRbs was performed in \cite{besmer} as 
these contributions are not relevant in the small \tz region, 
as can be seen from eqs.(\ref{glC7MI2}-\ref{glC8pMI2}). 

\begin{figure}[t]
\begin{tabular}{cc}
\hspace{-0.5cm}
\epsfig{file=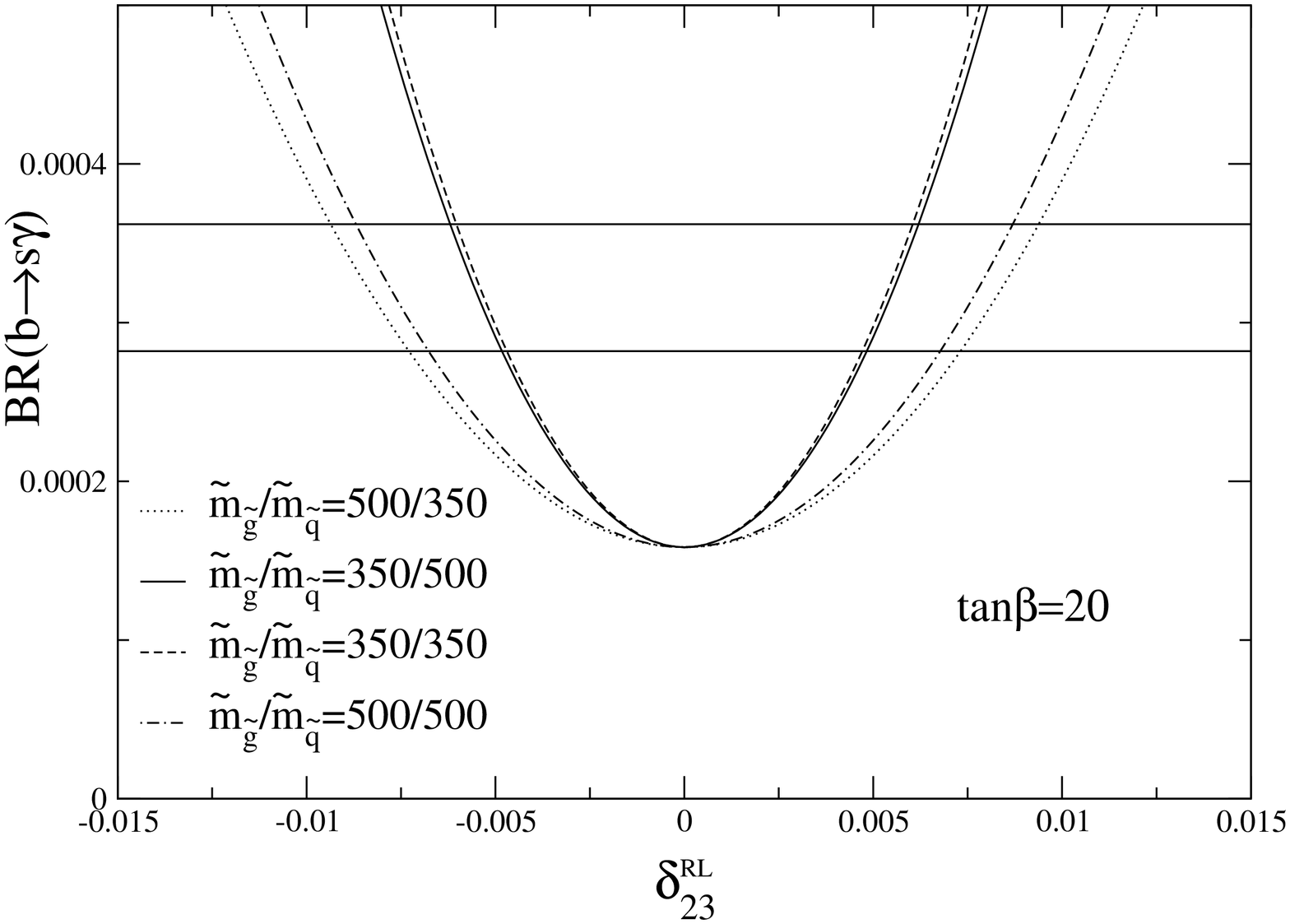,width=6.25cm,angle=-90} &
\hspace{-0.5cm}
\epsfig{file=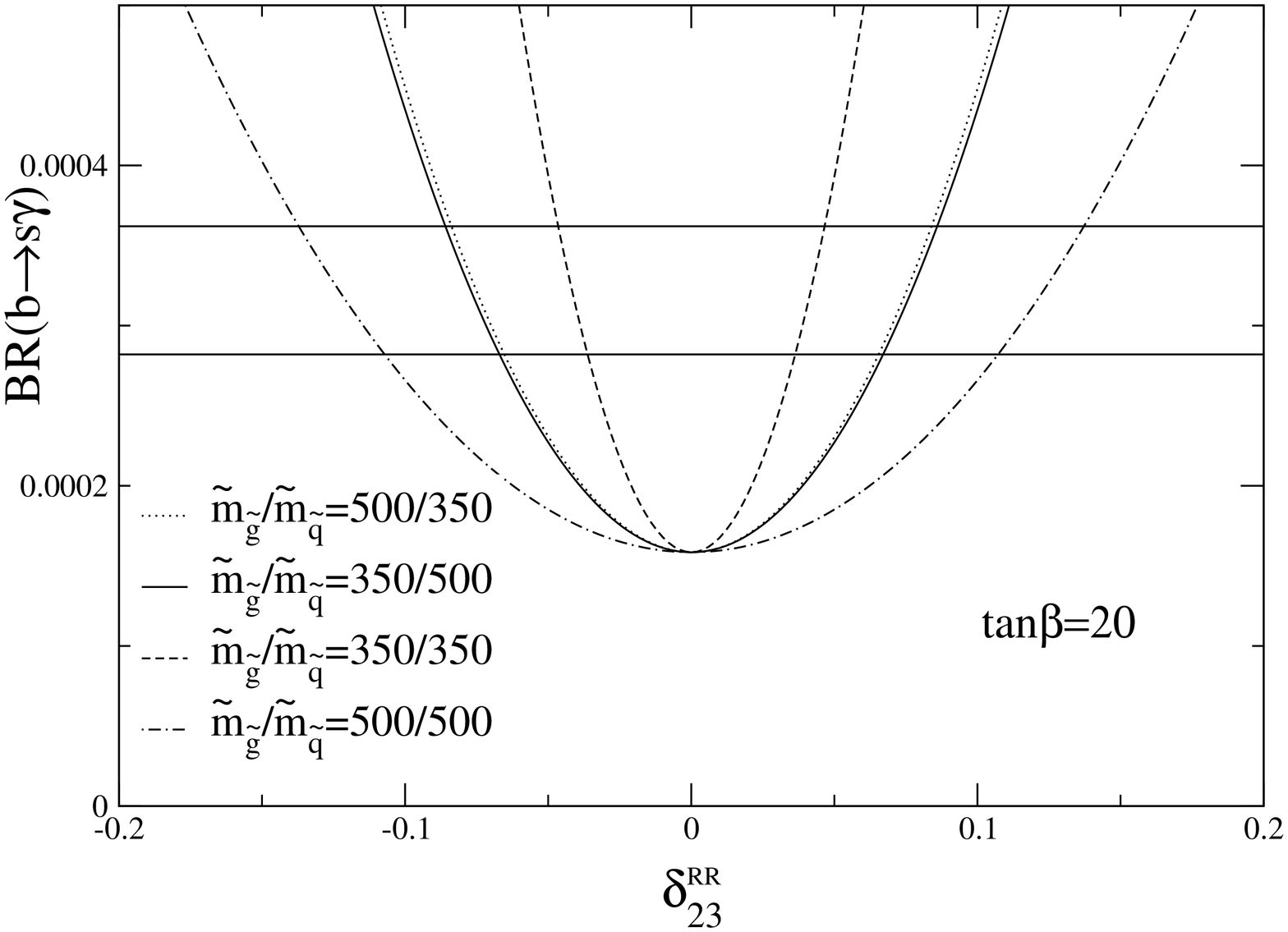,width=6.25cm,angle=-90} 
\end{tabular}
\caption{\it{The dependence of \bsg branching ratio on $\delta^{RL}_{23}$ 
and $\delta^{RR}_{23}$ for different values of $\tm_{\tilde{g}}/\tm_{D}$, 
for $\tz = 20$ and $\mu = 350$ GeV. All of the other off-diagonal entries 
except the one displayed in the axes, are assumed to vanish. 
$C_{7,8}(\mu_W) = 0$ is assumed. The horizontal lines represent the $1 
\sigma$ experimental allowed region.}}
\label{fig1}
\end{figure}
%

In Fig.~\ref{fig1} we show the dependence of the \bsg branching ratio 
on the MI terms \dRLbs and \dRRbs for different values of $x^g_D = 
\tm^2_{\tilde{g}}/\tm^2_{D}$ and for $\tz = 20$ and $\mu = 350$ GeV.   
All the other off-diagonal entries in the down-squark mass matrix are 
assumed to vanish for simplicity. ``Individual'' limits \dRLbs $< 10^{-2}$ 
and \dRRbs $< 1.5 \times 10^{-1}$ can be obtained respectively from the left 
and right side plot of Fig.~\ref{fig1}. Horizontal full lines represent 
$1 \sigma$ deviations from the experimental results reported in 
eq.(\ref{bsgex}). Of course, the required cancellation of the total 
$C_{7,8}$ contribution may in general need nonvanishing 
off-diagonal entries of the up and down squark mass matrices. However, the 
specific values of these entries do not significantly affect the absolute 
limits on \dRLbs and \dRRbs MIs shown in Fig.~\ref{fig1}. 

\begin{figure}[t]
\centerline{
\epsfig{file=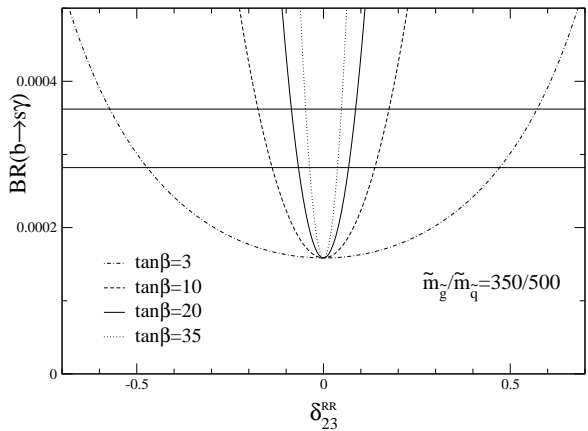,width=6.25cm, angle=-90} }
\caption{\it{Dependence of \bsg branching ratio on $\delta^{RR}_{23}$ 
for different three values of \tz, with the other parameters fixed to 
$\tm_{\tilde{g}}/\tm_{\tilde{q}} = 350/500$ and $\mu = 350$ GeV. All of 
the other off-diagonal entries, except the one displayed in the axes, 
are assumed to vanish. The horizontal lines represent the $1 \sigma$ 
experimental allowed region.}}
\label{fig2}
\end{figure}

As expected from eqs.(\ref{glC7pMI2},\ref{glC8pMI2}), the bounds obtained 
for \dRRbs are strongly dependent on the product $\mu \tan \beta$. In 
Fig.~\ref{fig2} we show the \tz dependence of this limit, for fixed 
$\tm_{\tilde{g}}/\tm_{\tilde{q}}=350/500$ and $\mu = 350$ GeV. 
More stringent bounds on \dRRbs can be obtained for larger \tz. For 
\tz$>35$ the bounds on \dRRbs can become as stringent as the \dRLbs 
bounds. 
Similar considerations and bounds obviously hold also for the \dLLbs MI. 
As we are only interested here in the gluino contributions to $C'_{7,8}$, 
we do not discuss this sector in detail. Clearly this term must be 
taken into consideration if a similar analysis was performed for the 
$C_{7,8}$ coefficient in the large \tz region.
%

\begin{figure}[t]
\vspace{0.1cm}
\begin{tabular}{cc}
\hspace{-0.5cm}
\epsfig{file=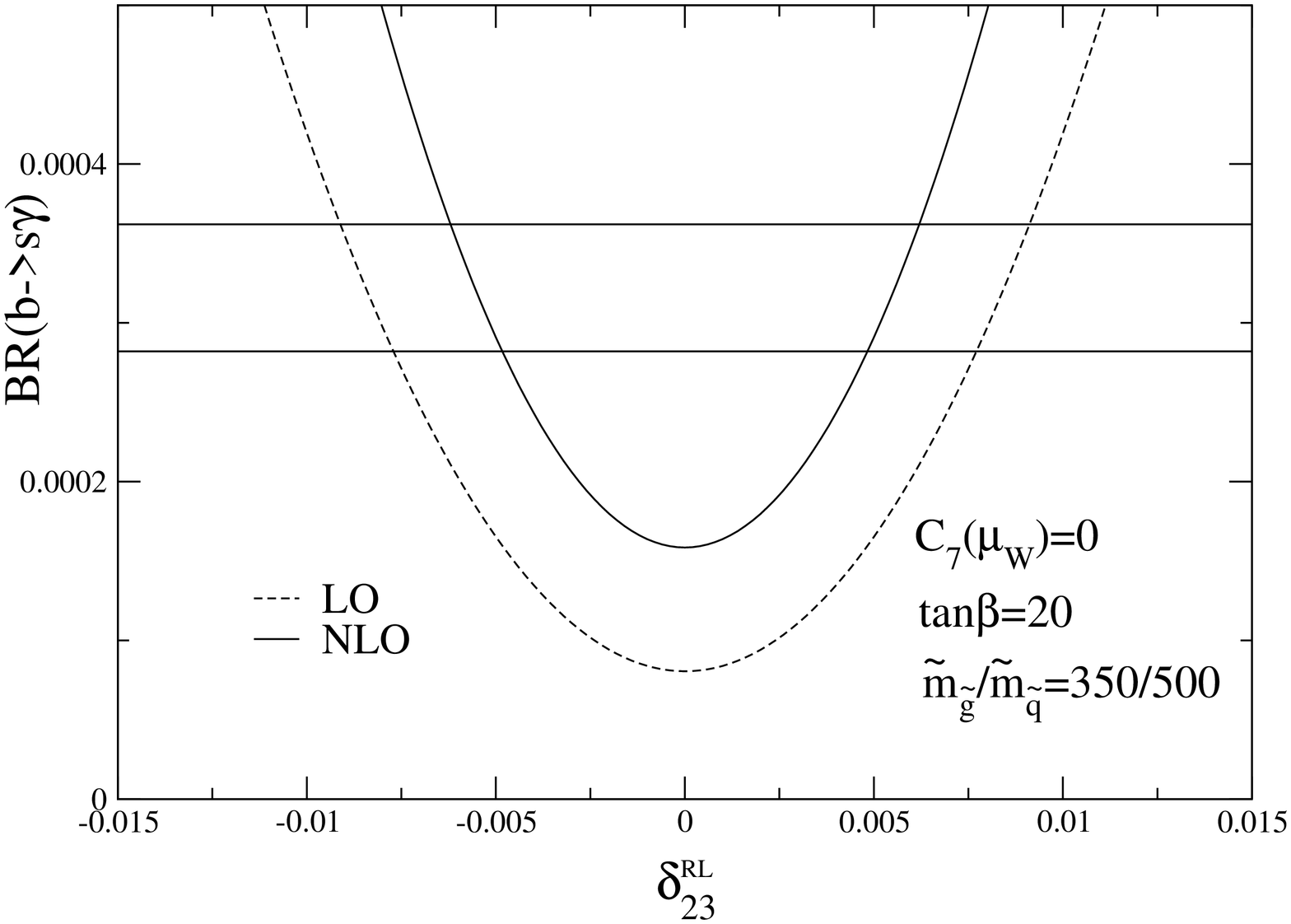, height=8.15cm, angle=-90} & 
\hspace{-0.5cm}
\epsfig{file=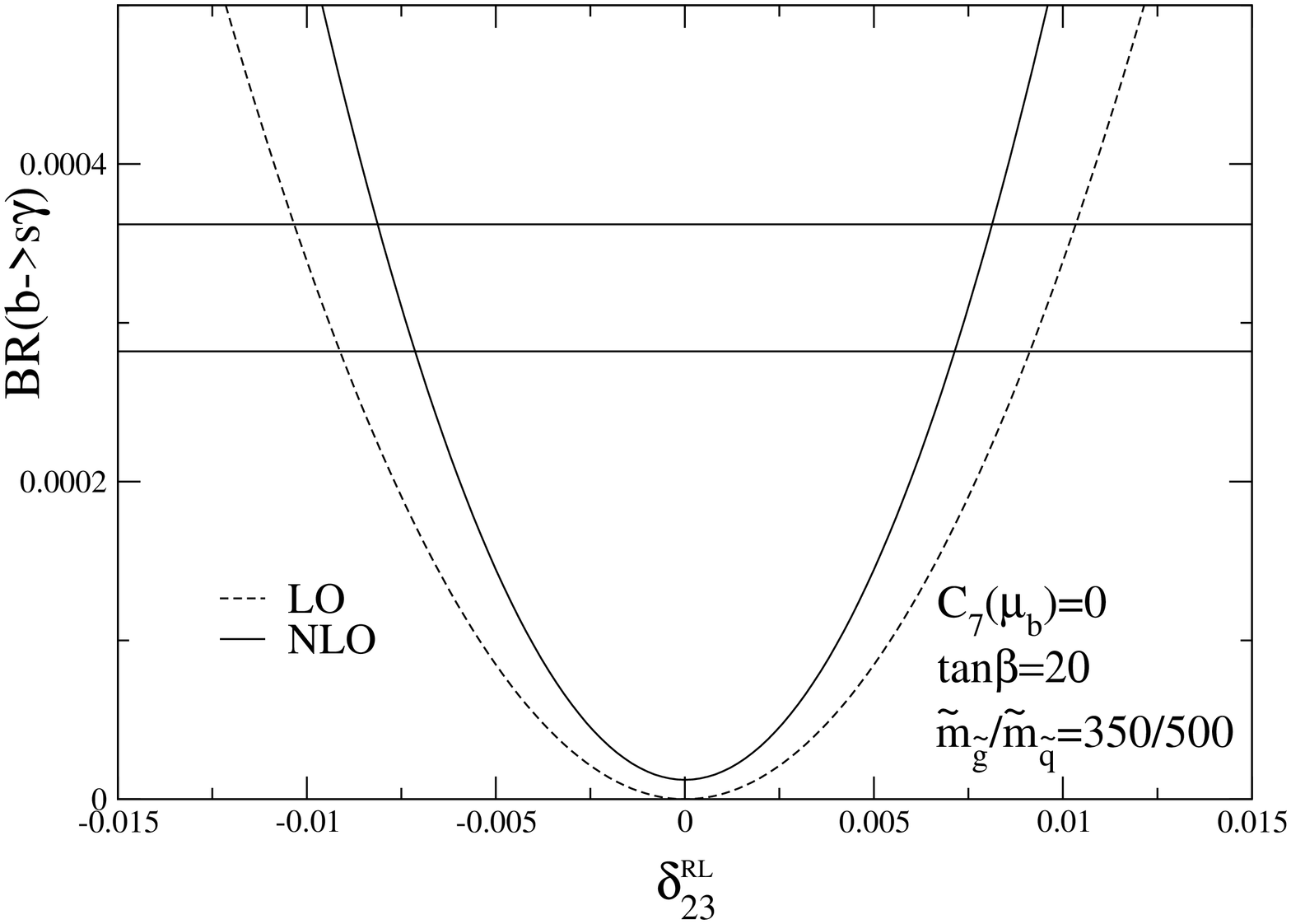, height=8.15cm, angle=-90} 
\end{tabular}
\caption{\it{Dependence of \bsg branching ratio on $\delta^{RL}_{23}$ 
for $\tm_{\tilde{g}}/\tm_{\tilde{q}}=350/500$, \tz$=20$ and $\mu=350$ 
GeV. All the other off-diagonal entries, except the one displayed in the 
axes, are assumed to vanish. In the plots we show the result 
obtained using LO (dashed line) and NLO (full line) formula for the 
\bsg branching ratio, setting respectively $C_{7,8}(\mu_W)=0$ (left plot) 
and $C_{7,8}(\mu_b)=0$ (right plot). The horizontal lines represent the $1 
\sigma$ experimental allowed region.}}
\label{fig3}
\end{figure}

In Fig.~\ref{fig1} and Fig.~\ref{fig2}, we set $C_7 = C_8 
= 0$ so that the only contribution to the \bsg branching ratio is due to  
the gluino contribution to $C'_7$ and $C'_8$. Thus one should think 
that for vanishing \dRLbs and/or \dRRbs the branching ratio in our scenario 
should vanish. The reason for the finite, nonzero contribution is the fact 
that we are using a NLO formula for the \bsg branching ratio \cite{kagan}. 
At NLO, imposing the condition $C_{7,8}(\mu_W)=0$ still leaves constant 
terms that arise from the mixing of the SM operators 
(specifically, in our chosen basis, $C_2$) that do not contribute to the 
branching ratio at LO. In Fig.~\ref{fig3} (left side), we compare the 
results obtained using the LO and NLO expression for the \bsg branching 
ratio imposing the condition $C_{7,8}(\mu_W)=0$. As can be seen 
explicitly, the difference in using the LO or NLO is sizeable. In 
Fig.~\ref{fig3} (right side), 
we compare the results obtained using the LO and NLO expression for the 
\bsg branching ratio imposing the condition 
$C_{7,8}(\mu_b)=0$. 
As one can see now, the LO contribution to the \bsg branching ratio 
vanishes for vanishing MIs. This does not happen for the LO contribution 
of the left plot, as a finite contribution to the branching ratio appears 
from the running $\mu_W \raw \mu_b$ when the condition $C_{7,8}(\mu_W)=0$ 
is taken. In all the plots, except Fig.~\ref{fig3} (right side), we 
use $C_{7,8}(\mu_W)=0$, as this is the natural scale where cancellations 
could be explained in terms 
of the underlying fundamental theory, while the choice $C_{7,8}(\mu_b)=0$ 
seems highly accidental. Finally, it should be noted that the strongest 
restriction comes from imposing the condition $C_{7}=0$. The same 
requirement on $C_8$ could easily be relaxed, and our results 
would remain almost unchanged. The $C_8$ contribution to the \bsg 
branching ratio represents in fact only a $10\%$ effect of the total.  

%
%

It is important to notice at this point that a consistent analysis of 
\bsg at NLO would require the calculation of the two-loop (QCD and SQCD) 
contribution to the $C'_7$ coefficient. In the general uMSSM the 
calculation of the ${\cal{O}}(\alpha_s^2)$ contribution to $C'_7$ (and 
obviously $C_7$) is extremely complicated. In \cite{bobeth}, the 
contribution to $C_7$ from the two-loop diagrams with one gluino and one 
gluon internal line 
has been calculated. This represents the dominant MSSM two-loop 
contribution only in the limit of very heavy gluino mass of ${\cal{O}}(1 
{\rm TeV})$ 
and small \tz ($\approx 1$). Thus it cannot be applied to our analysis, in 
which SUSY masses (and the gluino mass in particular) below $500$ GeV 
and large \tz are assumed. In fact, if the gluino mass is light  
the two-loop diagrams with two gluino internal lines should also be taken 
into account. Moreover, if \tz is large, diagrams with internal 
Higgsino lines cannot be neglected anymore as Yukawa couplings can 
become of ${\cal O}(1)$. Using the results of \cite{bobeth}, one obtains 
an effect of a few percent in the \bsg branching ratio. It should be 
remembered, however, that in our analysis this provides only a very crude 
estimation. It seems reasonable to expect a possible $10\%$ modification 
of the \bsg branching ratio results from the inclusion of the complete NLO 
calculation of the $C'_7$ coefficient. Moreover, while the two-loop 
diagrams with gluino/gluon internal lines have the same MI structure and 
as such are proportional to the one-loop gluino contribution to $C'_7$, 
this is not the case for the diagrams with gluino/Higgsino internal 
lines, for which the CKM flavor changing structure also enters.   
%
%
\subsection{General MI analysis}
\label{sectiondb}
A general analysis of the gluino contribution to $C'_{7,8}$ depends 
simultaneously on both the \dRLbs and \dRRbs MIs. For a complete 
specification of our scenario the only other free 
parameters that need to be fixed are the ratio between the gluino mass and 
the common down-squark mass, $\tm_{\tilde{g}}/\tm_D$, the product $\mu$ 
\tz, and the relative phase between \dRLbs and \dRRbs. The influence in 
of all the other down-sector squark matrix off-diagonal entries and 
MSSM parameters in the $C'_{7,8}$ sector can safely be 
neglected\footnote{Of course all of the other off-diagonal entries of the  
down-squark and up-squark mass matrices as well all the other flavor 
conserving MSSM parameters enter in our analysis, as we assume to choose 
them in such a way that the condition $C_{7,8}=0$ is satisfied. However, 
as previously mentioned the detailed analysis of this condition will be 
discussed in a following paper \cite{ekrww2}.}. Thus, 
we can have a complete description in terms of only five free parameters 
of the \bsg phenomenology in our MSSM ``$C'_7$ dominated'' scenario. 

\begin{figure}[t]
\vspace{-0.75cm}
\centerline{
\epsfig{file=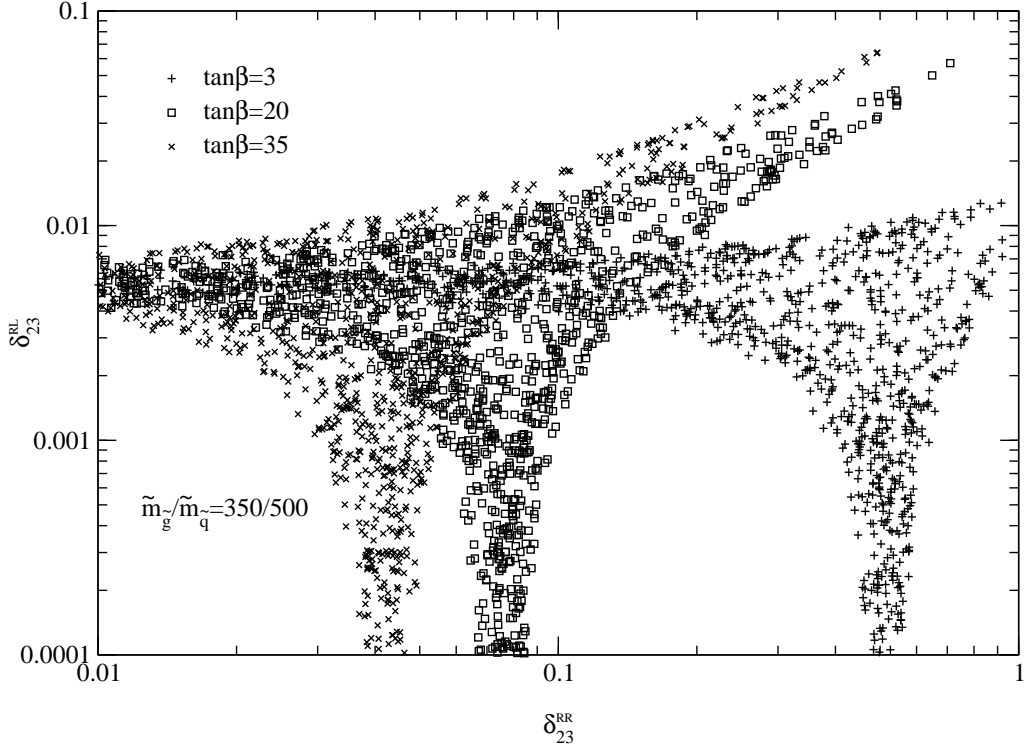, width=11cm, angle=-90} } 
\caption{\it{$1\sigma$-allowed region in the (\dRRbs$\!$, \dRLbs$\!$) 
parameter space for three different values of $\tan \beta$, with the 
other parameters fixed to $\tm_{\tilde{g}}/\tm_{\tilde{q}}=350/500$, and 
$\mu=350$ GeV. All the other off-diagonal entries, except the one 
displayed in the axes, are assumed to vanish.}}
\label{fig5}
\end{figure}

In Fig.~\ref{fig5} we show the $1 \sigma$ experimentally allowed region 
in the (\dRLbs$\!$, \dRRbs$\!$) parameter space for a specific choice of 
$\tm_{\tilde{g}}/\tm_D=350/500$, $\mu=350$ GeV, and for three different 
values of \tz=$3, 20$ and $35$. For \dRLbs or \dRRbs vanishing, one 
obtains the regions depicted in Figs.~\ref{fig1} and \ref{fig2}. 
Larger regions in the (\dRLbs$\!$, \dRRbs$\!$) parameter 
space are obtained when both the MIs take nonvanishing values. It is 
clear no absolute limit can be derived for the two MIs simultaneously. 
The values (\dRRbs$\!$, \dRRbs$\!$) $\approx (1,0.1)$ are, for example, 
possible\footnote{One should check if, for such large MI values, charge 
and color breaking minima appear. Anyway as these are usually rather 
model depend assumptions we don't introduce here the constraints discussed, 
for example, in \cite{ccb}.} for \tz$=35$. In fact, as can be seen in 
Fig.~\ref{fig5}, there is always a ``flat direction'' where large values 
of \dRLbs and \dRRbs can be tuned in such a way that the gluino 
contribution to $C'_{7,8}$ is consistent with the experimental bound. 
This flat direction clearly depends on the chosen values for 
$\tm_{\tilde{g}}/\tm_{\tilde{q}}$ and $\mu$\tz. 
The presence of this particular direction is explained by the fact that 
we are allowing complex off-diagonal entries. Hence the relative phase 
between \dRLbs and \dRRbs can be fixed in such a way that the needed 
amount of cancellation can be obtained between the first and second order 
MI contribution. In the notation used in eqs.(\ref{glC7pMI},\ref{glC7pMI2})
the line of maximal cancellation is obtained for $\varphi = \arg[$\dRLbs 
\dRRbs $] =\pm \pi$.  
%

%
\subsection{CP asymmetry and branching ratio}
\label{sectiondc}
In addition to the \bsg branching ratio, the experimental collaborations 
will provide in the following years more precise measurements of the 
\bsg CP asymmetry:
\be
{\cal A}_{CP}(\bsg) = \frac{BR(\bsg)-BR(\bsgb)}{BR(\bsg)+BR(\bsgb)}~.
\label{asyCP}
\ee 
The present best experimental value available \cite{cleo2} gives 
at $90\%$ CL level the following range:
\be
-0.27 < {\cal A}_{CP}(\bsg) < 0.10~,
\label{asyCPex}
\ee
which is still too imprecise for to provide useful tests for NP, although 
the measurement is expected to be upgraded soon.

The only flavor-violating and CP-violating source in the SM (and MFV 
scenarios) is given by the CKM matrix, which results in a very small 
prediction for the CP asymmetry. In the SM an asymmetry approximatively of 
$0.5 \%$ is expected \cite{kagan}. If other sources of CP violation are 
present, a much bigger CP asymmetry could be produced (see references 
\cite{kagan} and \cite{aoki}). 

\begin{figure}[t]
\centerline{
\epsfig{file=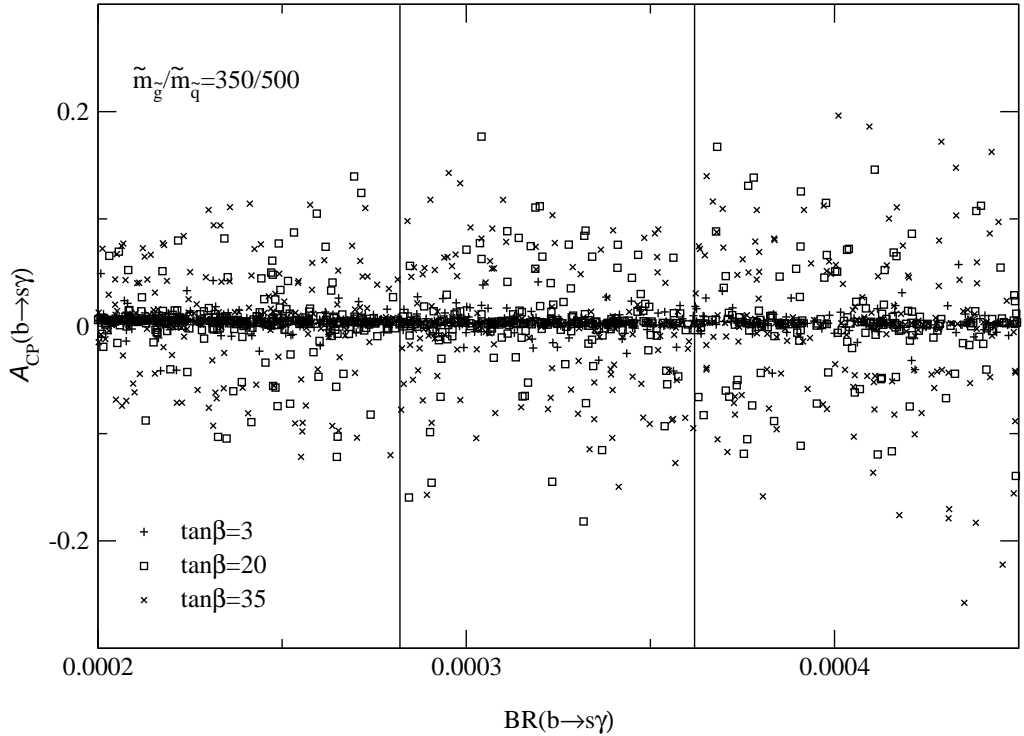, width=11cm, angle=-90} } 
\caption{\it{Asymmetry vs branching ratio for three different values of 
$\tan \beta$, with $\tm_{\tilde{g}}/\tm_{\tilde{q}}=350/500$, and 
$\mu=350$ GeV. All the off-diagonal entries except \dRLbs and \dRRbs are 
assumed to  vanish. The vertical lines represent the $1\sigma$ 
experimental allowed region.}}
\label{fig6}
\end{figure}
%


In our $C'_7$ dominated scenario, one can derive the following 
approximate relation for the CP asymmetry \cite{kagan}, in terms of 
the \dRLbs and \dRRbs MIs:
\be
{\cal A}_{CP}(\bsg \!) = -\frac{4}{9}\alpha_s(\mu_b) \frac{ 
   {\rm Im}\left[ C'_7 C'^*_8 \right]}{|C'_7|^2} \approx k(x^g_D)
   \left(\frac{m_b \mu \tz}{\tm^2_D} \right) 
   |\dRLbs \dRRbs \!\!| \sin \varphi~,
\label{asymMI}
\ee
in which $\varphi$ is the relative phase between \dRLbs and \dRRbs as 
previously defined. The constant of proportionality $k(x^g_D)$ 
depends only on the ratio $\tm_{\tilde{g}}/\tm_D$ through the integrals 
$F_i$ and can be easily obtained from eqs.(\ref{glC7pMI},\ref{glC7pMI2}). 
One can immediately note from eq.(\ref{asymMI}) that if only one MI is 
considered, the CP asymmetry is automatically zero.
Moreover, a nonvanishing phase in the off-diagonal down-squark mass 
matrix is necessary\footnote{Recall that for avoiding EDM constraints 
reparameterization invariant combinations of flavor-independent 
phases (such as the phase of $\mu$ in a particular basis) are taken to 
be zero.}. No sensitive bounds 
on this phase can be extracted from EDM's in a general flavor violating 
scenario. 

In Fig.~\ref{fig6}, we show the results obtained for the branching 
ratio and CP asymmetry in which \dRLbs$\!$, \dRRbs and 
the relative phase $\varphi$ are varied arbitrarily for a fixed value 
$\tm_{\tilde{g}}/
\tm_{\tilde{q}}=350/500$ and \tz$=35$. The full vertical lines 
represents the $1 \sigma$ region experimentally allowed by the 
\bsg branching ratio measurements. It is possible, using $C'_{7,8}$ 
alone, to saturate the \bsg measured branching ratio and at the same 
time have a CP asymmetry even larger than $\pm 10\%$, the sign of the 
asymmetry being determined by the sign of $\sin \varphi$. As 
Fig.~\ref{fig6} shows, in the relevant branching ratio range the 
CP asymmetry range is constant. No strong dependence from $\tan \beta$, 
in the large \tz region, is present. The points with large asymmetry 
($>5\%$) lie in the ``flat direction" observed in Fig.\ref{fig5} and 
they have almost $\varphi \approx \pm \pi$ (obviously for $\varphi = \pm 
\pi$ the CP asymmetry vanishes). The explanation of this fact is the 
following. The numerator is proportional to $\sin \varphi$ and so 
goes to 0 as $\varphi$ approaches $\pm \pi$. However, at the same time  
it is enhanced for large MI values. This happens when the flat 
direction condition is (almost) satisfied. Here, in fact, a cancellation 
between the two (large) MI terms takes place, providing the enhancement 
of the CP asymmetry as the denominator remains practically constant, fixed 
by the allowed experimental measurement on the branching ratio. Note also 
that for parameter values outside the flat direction condition a 
CP asymmetry of a few $\%$ can 
still be observed, about ten times bigger than the SM prediction. The 
same order of magnitude can be observed in MFV when large \tz effects are 
taken into account \cite{demir}. In our scenario even smaller values of 
the CP asymmetry can be 
obtained, e.g. if one of the two off-diagonal entries is negligible, or 
the two MIs are ``aligned''. 

%
\subsection{Distinguishing the ``$C'_7$ dominated'' scenario from the 
``$C_7$ dominated"}
\label{sectiondd}
A possible method for disentangling the relative contributions to the 
\bsg branching ratio from the $Q_7$ and $Q'_7$ operators utilizes an   
analysis of the photon polarization. A detailed analysis of how it
is possible to extract information from the photon polarization in 
radiative B decays is given in \cite{bsgpol}. For simplicity, let us 
define the following ``theoretical'' LR asymmetry at LO:
\be
{\cal A}_{LR}(\bsg \! \!) = 
   \frac{BR(b \raw s \gamma_L) - BR(b \raw s \gamma_R)}
        {BR(b \raw s \gamma_L) + BR(b \raw s \gamma_R)} = 
   \frac{|C_7(\mu_b)|^2-|C'_7(\mu_b)|^2}{|C_7(\mu_b)|^2+|C'_7(\mu_b)|^2}~,
\label{asyLR}
\ee
which could in principle disinguish between $C_7$ or $C'_7$ dominated 
scenarios. Here L,R is the polarization of the external photon. This 
quantity is related to the quark chiralities of the $Q_7, Q'_7$ operators. 
Note that the photon polarization is the best possibility to gain 
information on the operator chirality, which gets almost lost in $b$ and 
$s$ quark hadronization into spin zero mesons (in principle if hadronization 
into spin one states could be isolated, perhaps some information could be 
obtained). Such a measurement is not yet available as only the average 
quantity $BR(b \raw s \gamma_L) + BR(b \raw s \gamma_R)$ is reported 
experimentally. 

In the SM case, and in general in all the MFV and mSUGRA scenarios, only 
the $C_7$ coefficient gives a nonnegligible contribution to the \bsg 
branching ratio. Only the right-handed bottom quark (in the center of 
mass reference frame) can decay, producing a photon with Left polarization and 
${\cal A}_{LR}(\bsg \!\!) = 1$. Small deviations from unity are possible 
due to subleading $m_s/m_b$ terms and hadronization effects. In our 
scenario, where the $C_7$ contribution is negligible, only left-handed 
bottom quarks can decay, emitting a photon with Right polarization, 
which in turn predict ${\cal A}_{LR}(\bsg\!\!) = -1$. In any other MSSM 
scenario, with nonminimal flavor violation, any LR asymmetry between $1$ 
and $-1$ is allowed. Consequently, a measurement of $A_{LR}(\bsg \!\!)$ 
different from $1$ will be a clear indication of physics beyond the SM 
with a nonminimal flavor structure. It will be very interesting to know 
if (and how precisely) CLEO, BABAR, and BELLE can measure the LR asymmetry 
of eq.(\ref{asyLR}). 
%
%
\section{Conclusions}
%
%
In this letter, we have discussed an alternative explanation of the \bsg 
branching ratio in the MSSM with a nonminimal flavor structure. We 
analyzed in particular the gluino contribution to the Wilson coefficient 
$C'_7$ associated with the ``wrong'' chirality operator $Q'_7$. We show 
that this coefficient arises mainly from two off-diagonal entries: 
\dRLbs and \dRRbs. For scenarios in which  where the $C_{7,8}$ 
contributions to \bsg are small, (i.e. for regions in the MSSM parameter 
space where W, Higgs, chargino and gluino contributions to $C_{7,8}$ tend 
to cancel each other) $C'_{7,8}$ provides the dominant effect. 
We derived absolute bounds separately on each of these coefficients. We
then described the allowed region of (\dRLbs$\!$, \dRRbs$\!$) parameter 
space, as a function of $\tan 
\beta$. We observed that (for a fixed ratio 
$\tm_{\tilde{g}}/\tm_{\tilde{q}}$ and for each chosen value of $\mu \tan 
\beta$), there exists a ``flat direction'' where large (even 
${\cal O}(1)$) off-diagonal entries are allowed. Along this direction the 
relative phase between the two MI elements is $\varphi \approx \pi$. For 
the majority of parameter space in this scenario the CP asymmetry is less 
than $5\%$. Asymmetries as big as $20\%$ can be obtained along the 
``flat directions''. Finally, we suggested a possible quantity (a LR 
asymmetry) that (if measured) can help to disentangle the $C_7$ from the 
$C'_7$ contribution to the \bsg branching ratio. Any ${\cal A}_{LR}(\bsg) 
\neq 1$ would be an irrefutable proof of physics beyond the SM. In 
addition, in the framework of the general MSSM, 
it would indicate the existence of nonminimal flavor violation produced 
by off-diagonal entries in the down-squark mass matrix, generally related 
to a nonzero gluino contribution. In our ``$C'_7$ dominated'' scenario, 
where the gluino contribution produce the only ``visible'' effect, we 
obtain in particular the extreme value ${\cal A}_{LR}(\bsg) = -1$. It 
would be very interesting if such a quantity could be measured. One 
implication of our analysis is that previous results on MSSM parameters, 
including constraints on the ``sign of $\mu$'' (i.e. its phase relative to 
$A_t$), are more model dependent than have been generally assumed.
%
%
\section*{Acknowledgements}
We thanks A. Donini, D. Demir, F. Feruglio, B. Gavela and A. Masiero 
for reading the manuscript and for the useful comments provided. L.E., 
G.K. and S.R. thanks the Aspen Center for Physics for the warm 
hospitality and the very nice atmosphere offered during the final 
stage of this work. 
%
%


\end{document}